\documentclass[aps,manuscript,showpacs,showkeys,superscriptaddress]{revtex4-1}
\usepackage{graphicx}
\usepackage{epsfig}  
\usepackage{epsf}    
\usepackage{dcolumn}
\usepackage{bm}
\usepackage{dcolumn}
\usepackage{textcomp}
\usepackage[tbtags]{amsmath}
\usepackage{amsfonts}
\usepackage{float}
\usepackage{subfig}
\usepackage[]{hyperref}
  \hypersetup{
  unicode=false,          
  pdftoolbar=true,        
  pdfmenubar=true,        
  pdffitwindow=true,     
  pdfstartview={FitH},    
  pdfsubject={Getting the best out of T2K and NOvA},   
  pdfnewwindow=true,      
  pdfcreator={RevTeX},
  colorlinks=true,       
  linkcolor=red,          
  citecolor=blue,        
  urlcolor=blue,           
  }
\usepackage{hypcap}

\def\anu{{\bar\nu}}

\newcommand{\beq}{\begin{equation}}
\newcommand{\eeq}{\end{equation}}
\newcommand{\beqa}{\begin{eqnarray}}
\newcommand{\eeqa}{\end{eqnarray}}

\newcommand{\tx}{{\theta_{12}}}
\newcommand{\ty}{{\theta_{13}}}
\newcommand{\tz}{{\theta_{23}}}

\newcommand{\dl}{{\Delta_{31}}}
\newcommand{\ds}{{\Delta_{21}}}

\newcommand{\atil}{\hat{A}}
\newcommand{\dtil}{\hat{\Delta}}

\newcommand{\dcp}{\delta_{\mathrm{CP}}}
\newcommand{\nova}{NO$\nu$A~}
\newcommand{\pmue}{P(\nu_\mu \rightarrow \nu_e)}
\newcommand{\pmumu}{P(\nu_\mu \rightarrow \nu_\mu)}

\newcommand{\pmuebar}{P(\bar{\nu}_{\mu} \rightarrow \bar{\nu}_e)}
\newcommand{\pmumubar}{P(\bar{\nu}_{\mu} \rightarrow \bar{\nu}_\mu)}

\newcommand{\dchsq}{\Delta\chi^2}

\begin{document}


\title{Matter vs. vacuum oscillations at long-baseline accelerator neutrino experiments}

\author{Suman Bharti}
\email[Email Address: ]{sbharti@phy.iitb.ac.in}
\affiliation{Department of Physics, Indian Institute of Technology Bombay,
Mumbai 400076, India}
\author{Ushak Rahaman}
\email[Email Address: ]{ushakr@uj.ac.za}
\affiliation{Centre for Astro-Particle Physics (CAPP) and Department of Physics, University of Johannesburg, PO Box 524, Auckland Park 2006, South Africa}
\author{S. Uma Sankar}
\email[Email Address: ]{uma@phy.iitb.ac.in}
\affiliation{Department of Physics, Indian Institute of Technology Bombay,
Mumbai 400076, India}
\date{\today}
\begin{abstract}
The neutrino oscillation probabilities at the long-baseline accelerator neutrino experiments are expected to be 
modified by matter effects. We search for evidence of such modification in the data of T2K and NO$\nu$A, by fitting 
the data to the hypothesis of (a) matter modified oscillations and (b) vacuum oscillations. We find that vacuum 
oscillations provide as good a fit to the data as matter modified oscillations. Even extended runs of T2K and NO$\nu$A,
with 5 years in neutrino mode $(5 \nu)$ and five years in anti-neutrino mode $(5 \bar{\nu})$, can {\bf not} make a 
$3~\sigma$ distinction between vacuum and matter modified oscillations. The proposed experiment DUNE, with 
neutrino and anti-neutrino runs of 5 years each $(5 \nu + 5 \bar{\nu})$, can rule out vacuum oscillations by itself 
at $5~\sigma$ if the hierarchy is normal. If the hierarchy is inverted, a $5~\sigma$ discrimination against vacuum 
oscillations requires the combination of $(5 \nu + 5 \bar{\nu})$ runs of T2K, \nova and DUNE. 
\end{abstract}
\pacs{14.60.Pq,14.60.Lm,13.15.+g}
\keywords{Neutrino Mass Hierarchy, Long-Baseline Experiments}
\maketitle

\section{Introduction}
Neutrino oscillations provide a signal for physics beyond standard model. Over the past 20 years, a large number of 
experiments have determined neutrino oscillation parameters to better and better precision. Of these, there are two
different types of oscillations, one driven by a smaller mass-squared difference (labelled $\Delta_{21}$ by convention)
and another driven by a larger mass-squared difference (labelled $\Delta_{31}$ by convention). The first hint of the 
oscillations, driven by the larger mass-squared difference $\Delta_{31}$, was noted in the deficit of upgoing atmospheric 
$\nu_\mu$/$\bar{\nu}_\mu$ events by the pioneering water Cerenkov detectors IMB \cite{Casper:1990ac,BeckerSzendy:1992hq}
and Kamiokande \cite{Hirata:1992ku,Fukuda:1994mc}. The atmospheric neutrino experiment Super-Kamiokande 
\cite{Fukuda:1998mi} and the long-baseline accelerator neutrino experiments MINOS \cite{Michael:2006rx}, 
T2K \cite{Abe:2013fuq} and \nova \cite{Adamson:2017qqn} observed the spectral distortions in the survival 
probabilities of $\nu_\mu$ and $\bar{\nu}_\mu$. Initial analysis of these distortions was done under the hypothesis 
of {\bf vacuum oscillations} and the values of the oscillation parameters $|\Delta_{31}|$ and $\sin 2 \theta_{23}$ were obtained. 
These experiments could not distinguish between positive and negative values of $\Delta_{31}$. The case of positive 
$\Delta_{31}$ is called normal hierarchy (NH) and that of negative $\Delta_{31}$ is called inverted hierarchy (IH).
For $\sin 2 \theta_{23} < 1$, the same value is realized by two different values of $\theta_{23}$, one $< 45^\circ$
in the lower octant (LO) and the other $> 45^\circ$ in the higher octant (HO).

Due to the propagation of the neutrinos through earth matter, it is expected that the oscillation probabilities
would be modified by matter effects. These matter effects are sensitive to the sign of $\Delta_{31}$ and their observation
can lead to a determination of this sign. For baselines less than $1000$ km, the matter effects lead to 
negligibly small changes in $\nu_\mu/ \bar{\nu}_\mu $ survival probabilities \cite{Gandhi:2007td}. Thus, the 
$\nu_\mu/\bar{\nu}_\mu$ disappearance data of accelerator neutrino experiments lead to essentially the same values 
of $|\Delta_{31}|$ and $\sin 2 \theta_{23}$ for the three cases: (a) vacuum oscillations, (b) matter oscillations with NH 
and (c) matter oscillations with IH. In the case of atmospheric neutrinos, the survival probabilities $P_{\mu \mu}$ and 
$P_{\bar{\mu} \bar{\mu}}$ are expected to undergo significant changes due to matter effects. However, at present 
Super-Kamiokande is able to make only a small distinction between them \cite{Abe:2017aap}. A number of studies 
\cite{Petcov:2005rv,Gandhi:2007td,Ghosh:2013mga,Ajmi:2015uda} explored the sensitivity of future atmospheric 
neutrino detectors to observe these matter effects and determine whether $\Delta_{31}$ is positive or negative 
(or whether hierearchy is NH or IH). 

Two current long-baseline accelerator neutrino experiments, T2K~\cite{Abe:2011ks} and 
\nova\cite{Ayres:2007tu}, are looking for evidence of the matter modification of oscillation probabilities. 
These experiments measure the survival probabilities, $\pmumu$ and $\pmumubar$, and the 
oscillation probabilities, $\pmue$ and $\pmuebar$. For baselines of the order of $1000$ km or less, the changes in 
survival probabilities are negligibly small but the changes in the oscillation probabilities are noticeable. Hence the 
ability of these two experiments to search for the matter effects comes essentially from the measurement of $\pmue$ 
and $\pmuebar$~\cite{Lipari:1999wy,Narayan:1999ck}. But these appearance probabilities are also sensitive 
to the unknown CP violating phase $\delta_{\rm CP}$. Given a set of data, three solutions are likely to occur 
\cite{Mena:2004sa, Prakash:2012az}: 
\begin{itemize}
\item
matter modified oscillations with NH and $\delta_{\rm CP}^1$,
\item
vacuum oscillations and $\delta_{\rm CP}^2$ and
\item
matter modified oscillations with IH and $\delta_{\rm CP}^3$.
\end{itemize}
For NO$\nu$A, the changes induced in $P_{\mu e}$ and in $P_{\bar{\mu} \bar{e}}$ due to matter effects 
are comparable to the changes induced when the value of $\dcp$ is changed by $90^\circ$ ~\cite{Bharti:2018eyj}.
Hence, the measured value of $\dcp$ depends significantly on the oscillation hypothesis that is used.  
For T2K experiment, the matter effects lead to a smaller change in the appearance probabilities. 
Hence the value of $\delta_{\rm CP}$ determined from T2K data is less sensitive to whether matter effects 
are included or not. Establishing CP violation in neutrino oscillations is one of the most 
important goals of both current and future long-baseline accelerator neutrino experiments. To achieve this goal, 
it is important to establish a distinction between vacuum oscillations and matter modified oscillations.

The matter effect is included in the neutrino evolution in the form of the Wolfenstein matter term~\cite{msw1}
\begin{equation}
 A ({\rm in~eV^2})=0.76\times 10^{-4} \rho ({\rm in~gm/cc})E ({\rm in~GeV}), 
\end{equation}
where $\rho$ is the density of matter and $E$ is the energy of neutrino. The presence of this matter term modifies 
both the mass-square differences and the mixing angles and hence the neutrino survival/oscillations probabilities. 
Matter effects play a crucial role in the solution for solar neutrino deficit~\cite{Mikheev:1986gs,Mikheev:1986wj, 
Bahcall:2004ut}. The existence of matter effects in oscillations driven by $\Delta_{21}$ is established at a significance 
better than $5~\sigma$~\cite{Fogli:2005cq}. However, as mentioned above, there is no evidence as yet for matter 
effects in the oscillations driven by $\Delta_{31}$. The existence of matter effects at this scale should be verified
independently, just as the oscillations by the two different mass-squared differences are established indepedently. 

The expression for matter modified $\pmue$ for T2K and \nova experiments is given 
by~\cite{Cervera:2000kp,Freund:2001pn}
\begin{eqnarray}
\pmue & = & 
\sin^2 2 \ty \sin^2 \tz\frac{\sin^2\dtil(1-\atil)}{(1-\atil)^2}+ \nonumber\\ 
& & \alpha \cos \ty \sin2\tx \sin 2\ty \sin 2\tz \cos(\dtil+\dcp)
\frac{\sin\dtil \atil}{\atil} \frac{\sin \dtil(1-\atil)}{1-\atil},
\label{pmue-exp}
\end{eqnarray}
where $\hat{\Delta} = 1.27 \Delta_{31}L/E$, $\hat{A} = A/\Delta_{31}$ and 
$\alpha = \Delta_{21}/\Delta_{31}$.
The expression for $\pmuebar$ is obtained by replacing $A$ by $-A$ and $\dcp$ by $-\dcp$. 
The corresponding probabilities for vacuum oscillations can be obtained by taking the limit $A \to 0$. 
We note that the two appearance probabilities depend not only on the unknown hierarchy but also on 
the two other unknowns $\dcp$ and the octant of $\tz$. Depending on the measured values of 
$P_{\mu e}$ and $P_{\bar{\mu} \bar{e}}$ it is possible to cancel the change induced 
by the matter effects by choosing a wrong value of $\dcp$ and/or $\tz$~\cite{Bharti:2018eyj}. 
Hence establishing unambiguous evidence for matter effects at long-baseline experiments is non-trivial.

In this work we study the distinction made by the present data of T2K and \nova between vacuum oscillations and 
matter modified oscillations (for both hierarchies). We find that matter modified oscillations with normal 
hierarchy do provide the best fit solution to the data but the vacuum oscillations provide nearly as good a fit. We 
also consider the ability of T2K and \nova to make a distinction by the end of their runs. We find that a $3~\sigma$ 
discrimination is not possible even if each experiment has a ($5\nu +5\bar\nu$) run (that is, $5$ years each in 
neutrino and anti-neutrino modes). We further study the ability of the future experiment DUNE to make such a 
discrimination. The baseline and hence the energy of the neutrino beam of DUNE are larger which lead to larger 
changes in $\pmue$ and $\pmuebar$ due to matter effects. A one year neutrino run of DUNE by itself can make a 
$3~\sigma$ discrimination between matter and vacuum oscillations, if the hierarchy is normal. Addition of the T2K and 
\nova data to one year of DUNE data leads only to a small improvement in this discrimination. Vacuum oscillations can 
be ruled out at $5~\sigma$, for both normal and inverted hierarchies, by the combined T2K ($5\nu +5\bar\nu$), \nova 
($5\nu +5\bar\nu$) and DUNE ($5\nu +5\bar\nu$) runs. 

\section{Analysis procedure}
We use the following procedure to generate our results. We calculated the theoretical event spectra with 
three flavour oscillations using GLoBES~\cite{Huber:2004ka,Huber:2007ji}, for the appearance and 
disappearance channels in both neutrino and anti-neutrino modes for T2K and for NO$\nu$A. 
We have tuned the efficiencies in the software such that we get a match with the observed
number of events when the best-fit oscillation parameters are used as input.
These rates are calculated with the matter potential parametrized as $q*A$, where $A$ is the standard
Wolfenstein matter term
and $q$ is a multiplicative factor. In this analysis, we consider the possibility of non-standard matter term, as was 
done in ref.~\cite{Abe:2017aap}. 
The following inputs are used in our calculations: the solar neutrino parameters 
$\sin^2\tx$ and $\ds$ were held fixed at $0.31$ and $7.39 \times 10^{-5}\, {\rm eV}^2$ 
respectively. The values of $\sin^2 \ty$ were varied in its $3\, \sigma$ range around its
central value $0.02237$ ($0.02259$) with  $\sigma = 0.00066$ $(0.00065)$ for NH (IH). The values of $\sin^2 \tz$ were 
varied in its $3\, \sigma$ range
around its central value $0.563$ ($0.565$) with $\sigma = 0.024$ ($0.022$) for NH (IH). The value of
$|\dl|$ was varied in its $3\, \sigma$
range around its central value $2.528 \times 10^{-3}\, {\rm eV}^2$ ($2.510 \times 10^{-3}\, {\rm eV}^2$) with 
$\sigma=0.031 \times 10^{-3}\, {\rm eV}^2$ for NH (IH). The CP violating phase $\dcp$ is varied in its full range 
($0, 360^{\circ}$). The best-fit values and the $3~\sigma$ ranges of the measured neutrino oscillation
parameters are taken from ref. \cite{Esteban:2018azc, nufit}. The theoretical event rates are calculated separately 
for both the test hierarchies, NH and IH. The 
non-standard matter interaction parameter $q$ is varied between ($0,2$). The value $q=0$, of course, stands for 
vacuum oscillations.

In the first instance, we compare these theoretical event rates with the present data of T2K and NO$\nu$A. This is 
done by computing the $\chi^2$ between the theory and data for each of the four data sets of each experiment. 
For a particular experiment and for a particular data set,
the Poissonian $\chi^2$ is calculated using the expression
\begin{eqnarray}
\chi^2&=& \sum_i 2[{(N_i^{\rm th} - N_i^{\rm exp}) + N_i^{\rm exp}
\times \ln(N_i^{\rm exp}/N_i^{\rm th})}] + \sum_j [2\times N_j^{\rm th}]+\chi^2({\rm sys}),
\label{poisionian}
\end{eqnarray}
where $i$ stands for bins for which $N_i^{\rm exp}\neq 0$ and 
$j$ stands for bins for which $N_j^{\rm exp} = 0$. The term $\chi^2({\rm sys})$ arises due to systematic 
uncertainties. For each of the two experiments, we included systematic uncertainties
of $10\%$, using the pull method. We varied the pull parameter in $3\sigma$
range and marginalized over it to determine $\chi^{2}_{m}$ as a function of test values
of oscillation parameters, mass hierarchies and $q$. 
As explained in the introduction, the disappearance data in these experiments has negligible 
sensitivity to matter effects. The main sensitivity comes from the appearance data, which has a limited 
number of events because of the small value of $\theta_{13}$. The appearance events
are significant in a small number of energy bins. The sensitivity to matter effects
is {\bf more dependent} on the total number of apperance events than on the
energy distribution. The largest systematic uncertainty in this number is in the
number of expected $\nu_\mu$ events in the case of {\it no oscillations}. We 
took this systematic uncertainty to be $10\%$ and neglected the small variation
across the small number of energy bins.

We calculated the total $\chi^2$ for both NH test and IH test as 
\begin{eqnarray}
\chi^2({\rm tot})&=&\chi^{2}_{m}({\rm NO}\nu{\rm A}\, \nu\, {\rm app})+ \chi^{2}_{m}({\rm NO}\nu{\rm A}\, \anu\, {\rm 
app})+\chi^{2}_{m}({\rm T2K}\, \nu\, {\rm app})\nonumber \\
& &+ \chi^{2}_{m}({\rm T2K}\, \anu\, {\rm app})
+ \chi^{2}_{m}({\rm NO}\nu{\rm A}\, \nu\, {\rm disapp})
+ \chi^{2}_{m}({\rm NO}\nu{\rm A}\, \anu\, {\rm disapp})\nonumber \\
& &+\chi^{2}_{m}({\rm T2K}\, \nu\, {\rm disapp})+ 
\chi^{2}_{m}({\rm T2K}\, \anu\, {\rm disapp})+\chi^2({\rm prior}).
\end{eqnarray}
We added priors on $\sin^2 \ty$, $\sin^2\tz$ and $|\dl|$. During the 
calculation of $\chi^2({\rm tot})$, we have 
to keep in mind that the test values of the oscillation parameters are the same for all the individual $\chi^2$s. 
This quantity $\chi^2({\rm tot})$
is a function of test values of oscillation parameters, hierarchies and $q$. We  found the minimum of 
$\chi^2({\rm tot})$
and subtracted it from all other values of $\chi^2({\rm tot})$ to obtain $\dchsq$ as a function of test values of 
oscillation 
parameters, hierarchies and $q$. At the last step, we marginalized $\dchsq$ over all the oscillations parameters but 
not over hierarchy and $q$. 

In later stages we simulated the expected data from the future runs of T2K and NO$\nu$A and also the runs of the 
future experiment DUNE. These simulations were done with the best fit values~\cite{Esteban:2018azc} of mass-squared 
differences, mixing angles, $\dcp$ and $q=1$. These were done separately for both NH and for IH as true hierarchy.   
The results of these simulations were used as data and the theoretical event rates and $\chi^2$ were calculated as 
described earlier. In the case of this calculation, $\chi^2$ is equivalent to $\dchsq$.

\section{Results}
\subsection{From present data}
So far \nova has taken data $8.85\times 10^{20}$ POT in $\nu$ mode and $12.33\times 10^{20}$ POT in $\bar\nu$ mode.
The disappearance and appearance spectra for both modes are given in ref.~\cite{Acero:2019ksn}.
At present, T2K has taken data with $14.9\times 10^{20}$ POT in $\nu$ mode and $16.4\times 10^{20}$ POT in 
$\bar\nu$ mode~\cite{Abe:2019vii}. The appearance event spectra (for both $\nu$ and $\bar\nu$ modes) are given in 
ref.~\cite{Abe:2019vii} but the disappearance event spectra for the full data set are not available. 
Ref.~\cite{Abe:2018wpn} gives the disappearance spectra for $14.7\times 10^{20}$ POT in $\nu$ mode and $7.6\times 
10^{20}$ POT in $\bar\nu$ mode. We use the above mentioned spectra in our analysis. As explained in the introduction, the 
change induced in the survival probabilities due to matter effects is quite small. Therefore, we expect that 
our results would remain the same even when the disappearance spectra of the full T2K data are used in the analysis.

\begin{figure}[t]
\centering
\includegraphics[width=1.0\textwidth]{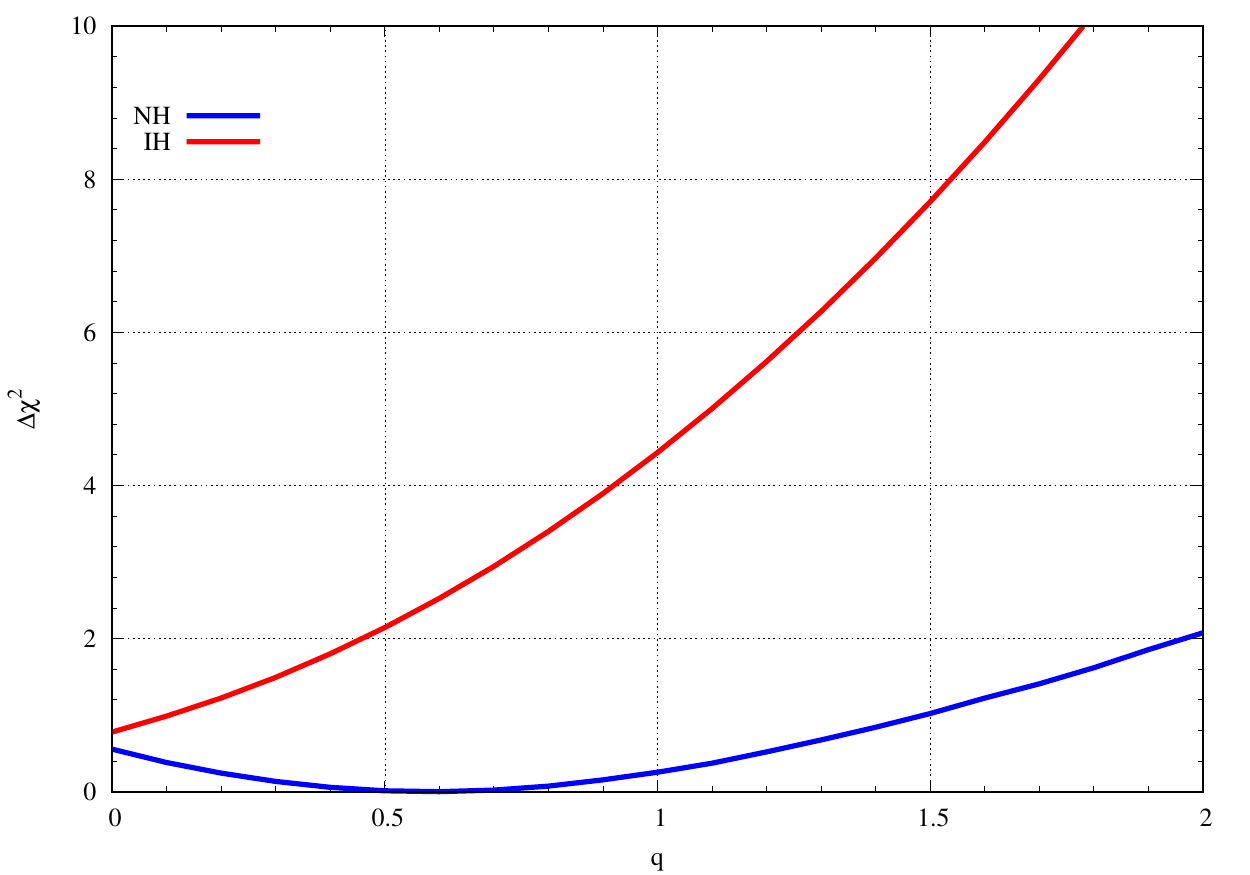}
\caption{\footnotesize{$\Delta \chi^2$ vs $q$ for the present data of T2K and NO$\nu$A. 
The data of \nova consists of $8.85\times 10^{20}$ POT in $\nu$ mode and $12.33\times 10^{20}$ POT 
in $\bar\nu$ mode. The appearance data of T2K consists of $14.9\times 10^{20}$ POT in $\nu$ mode 
and $16.4\times 10^{20}$ POT in $\bar\nu$ mode. The disappearance data of T2K consists of 
$14.7\times 10^{20}$ POT in $\nu$ mode and $7.6\times 10^{20}$ POT in $\bar\nu$ mode.  
The minimum $\chi^2$, for $152$ data points is $173.2$. The blue (red) curve is for NH (IH).
Note that vacuum oscillations $(q=0)$ also provide a good fit to the data.
}}
\label{fig1}
\end{figure}

The above data set, with $152$ data points, was fit to the hypothesis of three flavour oscillations with variable 
matter term as described in the previous section and the results are displayed in fig.~\ref{fig1}. The minimum 
$\chi^2=173.2$ occurs for $\Delta_{31}$ positive and $q=0.7$. Standard matter oscillations with NH ($q=1$) has 
essentially the same $\chi^2$ whereas the standard matter effects with IH are disfavored by a $\Delta\chi^2=4.5$. 
It is interesting to note that vacuum oscillations ($q=0$) provide nearly as good a fit to the data as matter modified 
oscillations with NH (with $\chi^2 = 173.8$). 

\begin{figure}[t]
\centering
\includegraphics[width=1.0\textwidth]{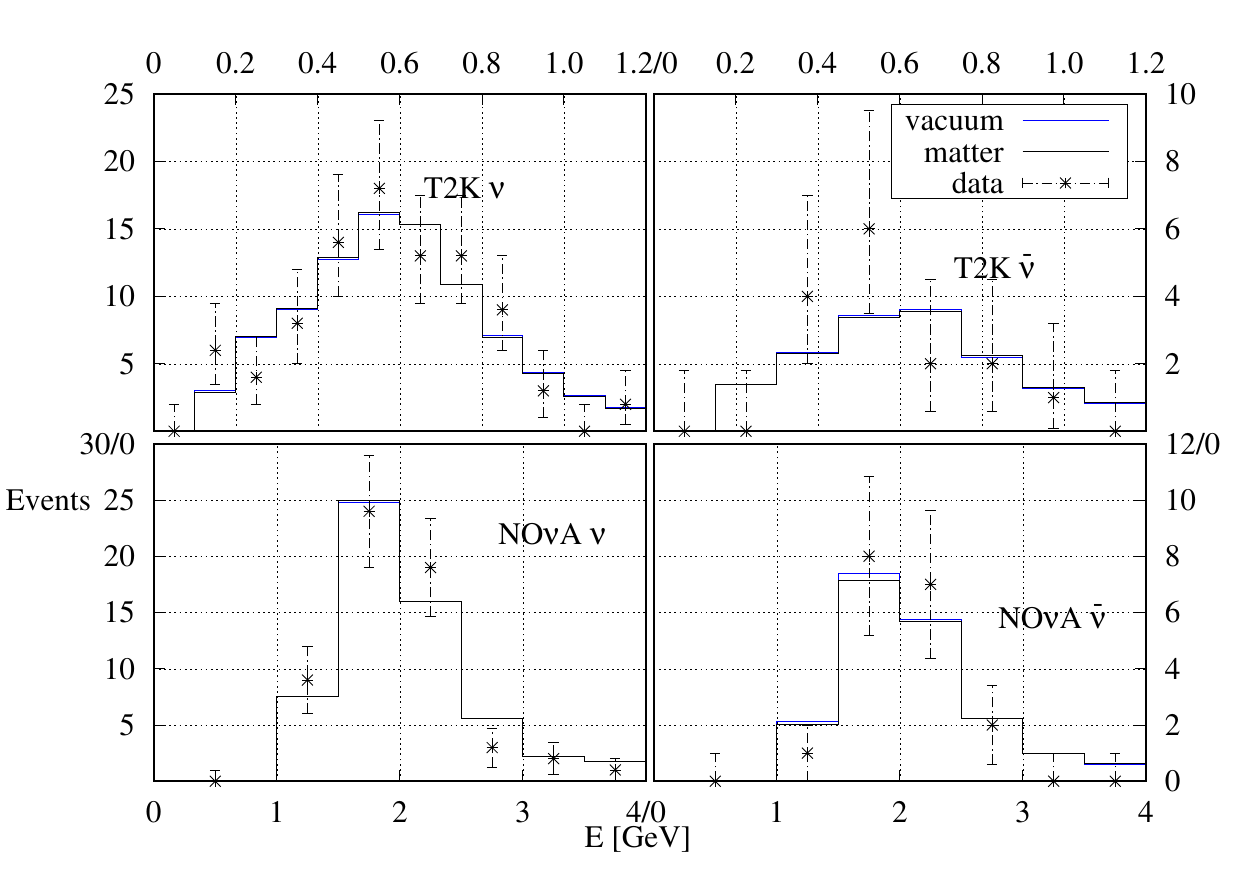}
\caption{\footnotesize{The observed number of events in appearance mode (shown as stars)
and the corresponding predictions from best-fit vacuum oscillations (blue histogram) and 
best-fit matter modified
oscillations (black histogram) for T2K (upper row) and \nova (lower row). The figures in the 
left panels are for $\nu_e$ appearance and those in the right panels are for $\bar{\nu}_e$ 
appearance.
}}
\label{figA}
\end{figure}

To explicitly verify that the best-fit vacuum oscillations provide nearly as good a fit to the data as the 
best-fit matter modified oscillations, we have plotted the observed appearance event numbers
in both neutrino and anti-neutrino modes for the two experiments, along with the predicted 
rates. The results are shown in fig.~\ref{figA}. From the plots in this figure, we note that there
is hardly any difference between the predictions of vacuum and matter modified oscillations 
and each fits the data very well.

\begin{figure}[t]
\centering
\includegraphics[width=1.0\textwidth]{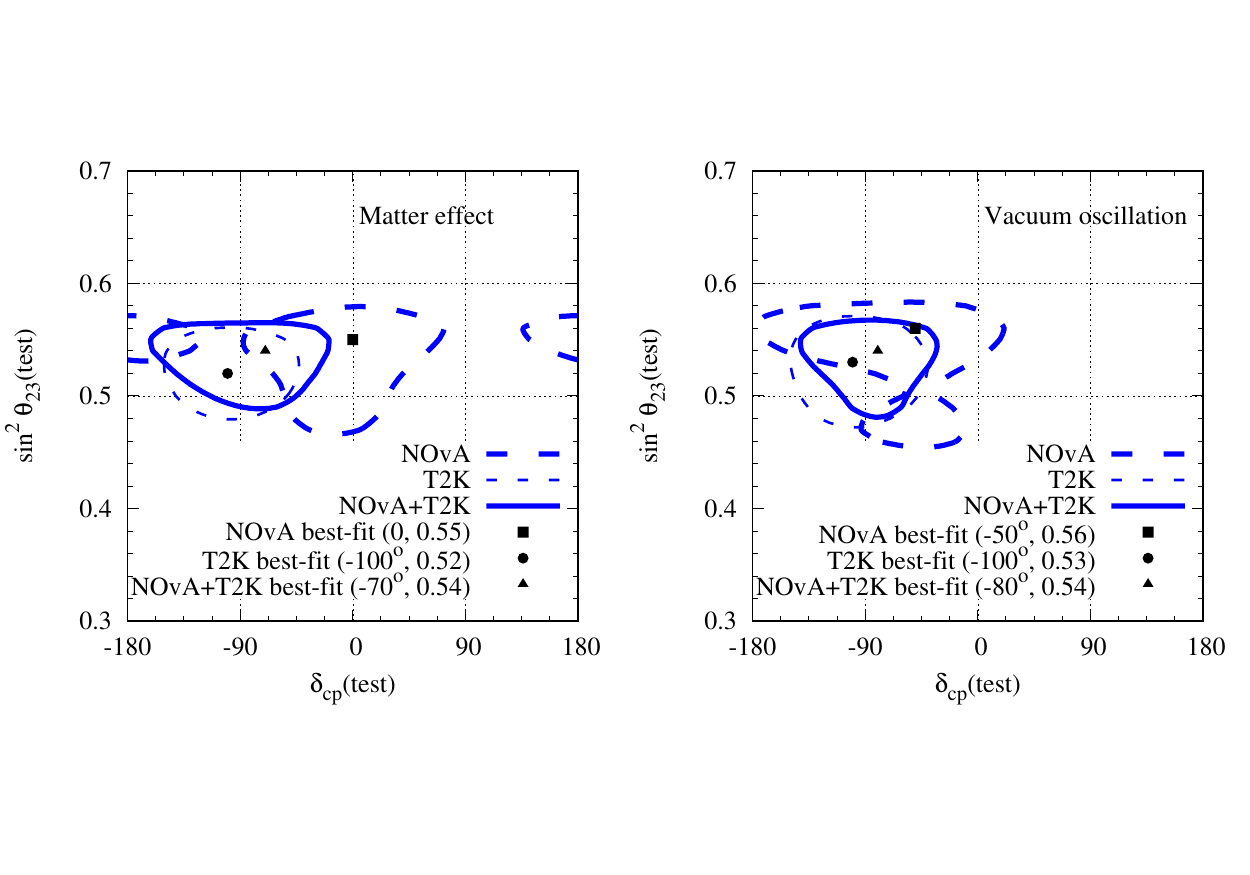}
\caption{\footnotesize{The best-fit points and $1~\sigma$ contours of T2K data (thin 
dashed lines), \nova data (thick dashed lines) and T2K+\nova data (solid lines) for
the case of matter modified oscillations (left panel) and vacuum oscillations (right panel).
}}
\label{figB}
\end{figure}

As stated in the introduction, the best-fit value of $\dcp$ depends on the oscillation hypothesis 
that is used to fit the data. In fig.~\ref{figB}, we show the best-fit points and $1~\sigma$ 
contours of T2K data, \nova data and T2K+\nova data for the two cases of matter modified
and vacuum oscillations. Comparing the two cases, we find that the best-fit point of T2K is
essentially the same for both vacuum and matter modified oscillations whereas the best-fit
value of $\dcp$ in the case of \nova strongly depends on the oscillation hypothesis used in the
fit. It is also interesting to note that there is {\bf less discrepancy} between the best-fit points
of the two experiments in the case of vacuum oscillations than in the case of matter modified
oscillations. 

\subsection{Expectation from extended runs of T2K and \nova }
\begin{figure}[t]
\centering
\includegraphics[width=1.0\textwidth]{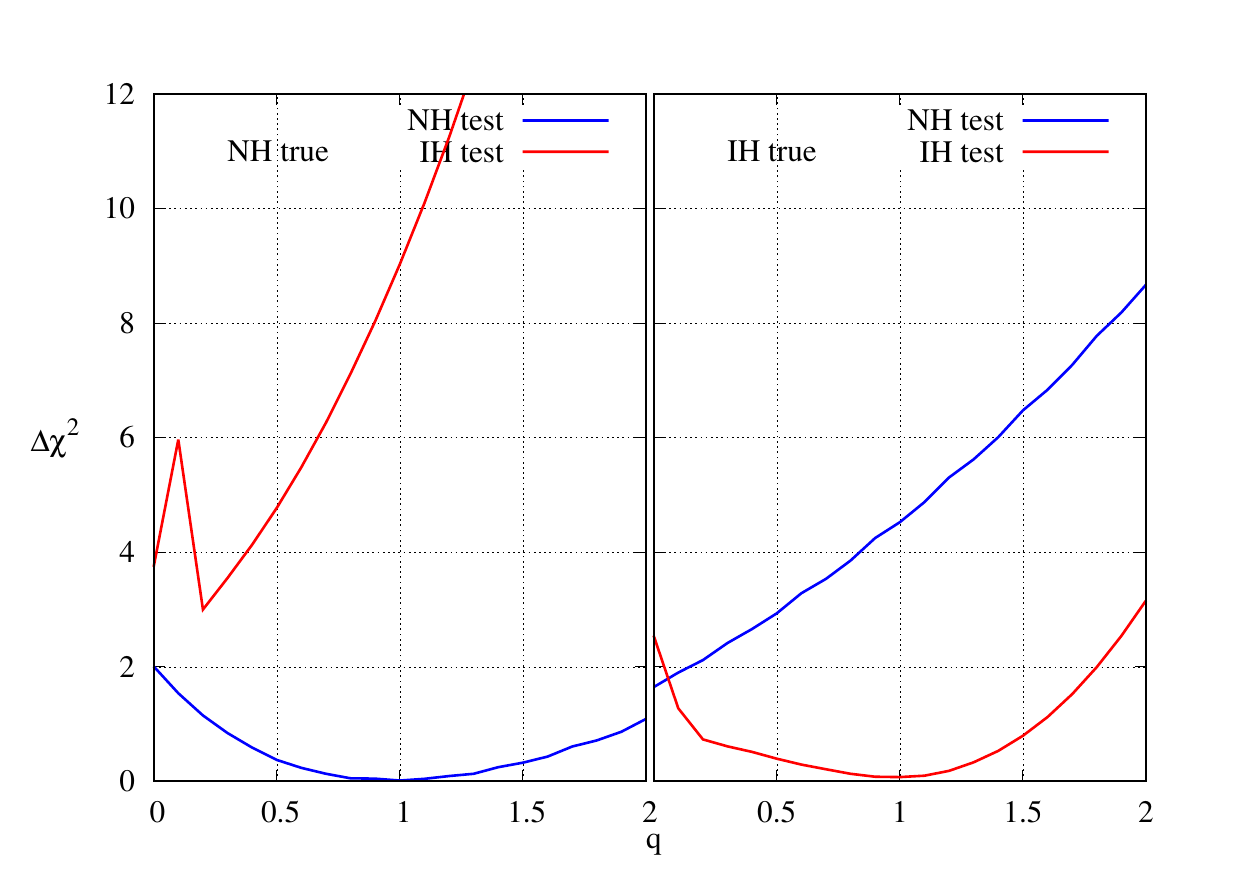}
\caption{\footnotesize{$\Delta \chi^2$ vs $q$ for an expected T2K run with $37.4\times 10^{20}$ POT each in both 
$\nu$ and $\bar\nu$ mode and a NO$\nu$A run with $30.25\times 10^{20}$ POT each in both $\nu$ and $\bar\nu$ mode,
equivalent to $(5 \nu + 5 \bar{\nu})$ runs for each experiment. The 
left (right) panel assumes the true hierarchy to be normal (inverted). The blue (red) curves assume the test 
hierarchy to be normal (inverted). Note that  data of T2K + \nova has {\bf no} ability to discriminate against vacuum oscillations 
even with such long exposure.
}}
\label{fig2}
\end{figure}

We now consider the ability of extended runs of T2K and \nova to establish matter modified oscillations. Using 
GLoBES, we simulated event spectra of T2K for $37.4\times 10^{20}$ POT each in both $\nu$ and 
$\bar\nu$ mode corresponding to a five year run in each mode. The simulation for \nova was done for $30.25\times 
10^{20}$ POT each in both $\nu$ and $\bar\nu$ mode, again corresponding to a five year run in each mode. We used the 
best fit values, taken from the global fits~\cite{Esteban:2018azc, nufit}, for the neutrino oscillation parameters in doing these
simulations, which were done separately for NH being the true hierarchy and for IH being the true hierarchy. The 
simulated 
data is analyzed in the manner described in section 2 and the results are shown in fig.~\ref{fig2}. We note that such 
an extended run rules out IH at $3~\sigma$ if NH is the true hierarchy but rules out NH only at $2~\sigma$ if IH is 
the true hierarchy. We also note that, no matter what the true hierarchy is, the vacuum oscillations have a very 
small $\Delta\chi^2\simeq 2$. Therefore the combined data of T2K and \nova {\bf cannot} distinguish 
between matter modified oscillations and vacuum oscillations.   

\subsection{Expectation from DUNE}
\begin{figure}[t]
\centering
\includegraphics[width=1.0\textwidth]{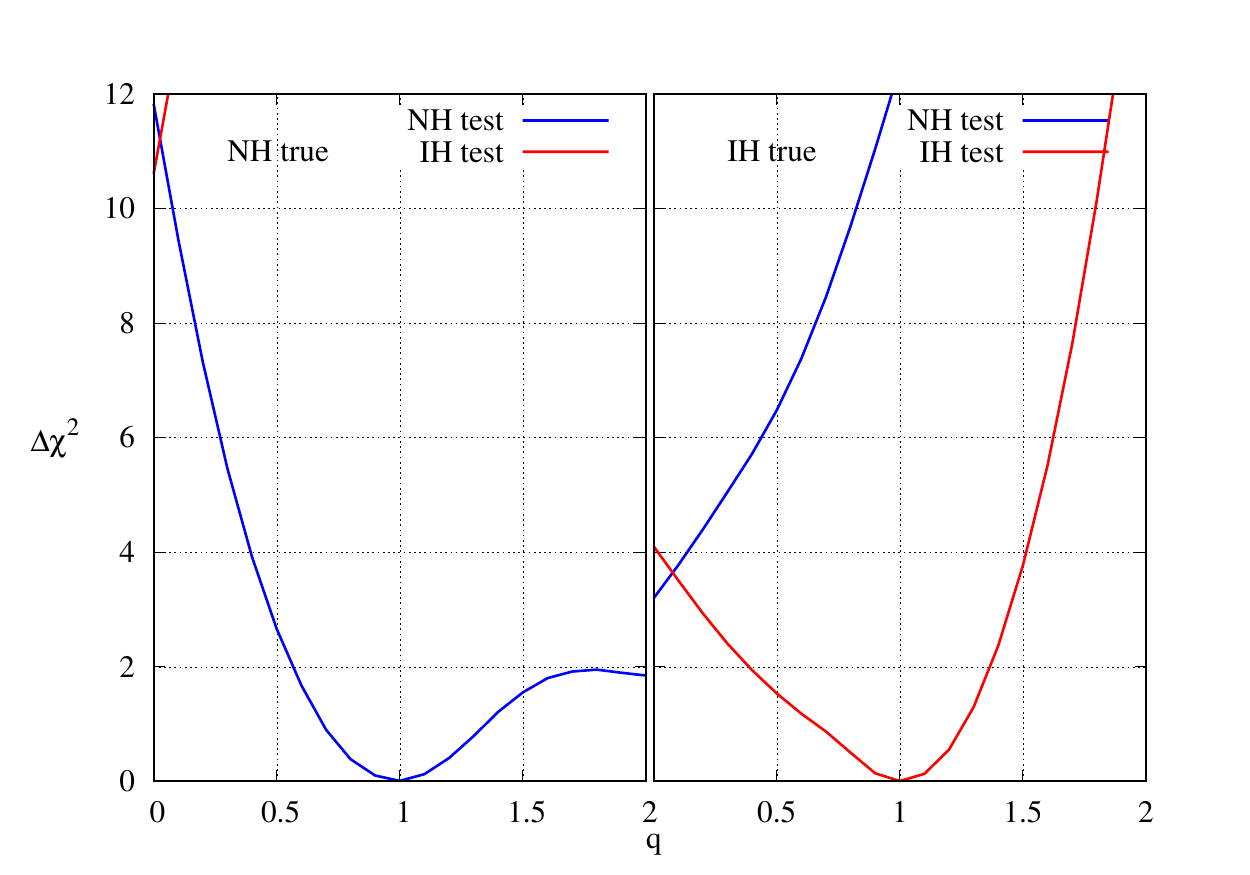}
\caption{\footnotesize{$\Delta \chi^2$ vs $q$ for an expected one year neutrino run of DUNE ($14.7 \times 10^{20}$ POT). 
The left (right) panel assumes the true hierarchy to be normal (inverted). The blue (red) curves assume the test 
hierarchy to be normal (inverted). 
}}
\label{fig3}
\end{figure}

\begin{figure}[t]
\centering
\includegraphics[width=1.0\textwidth]{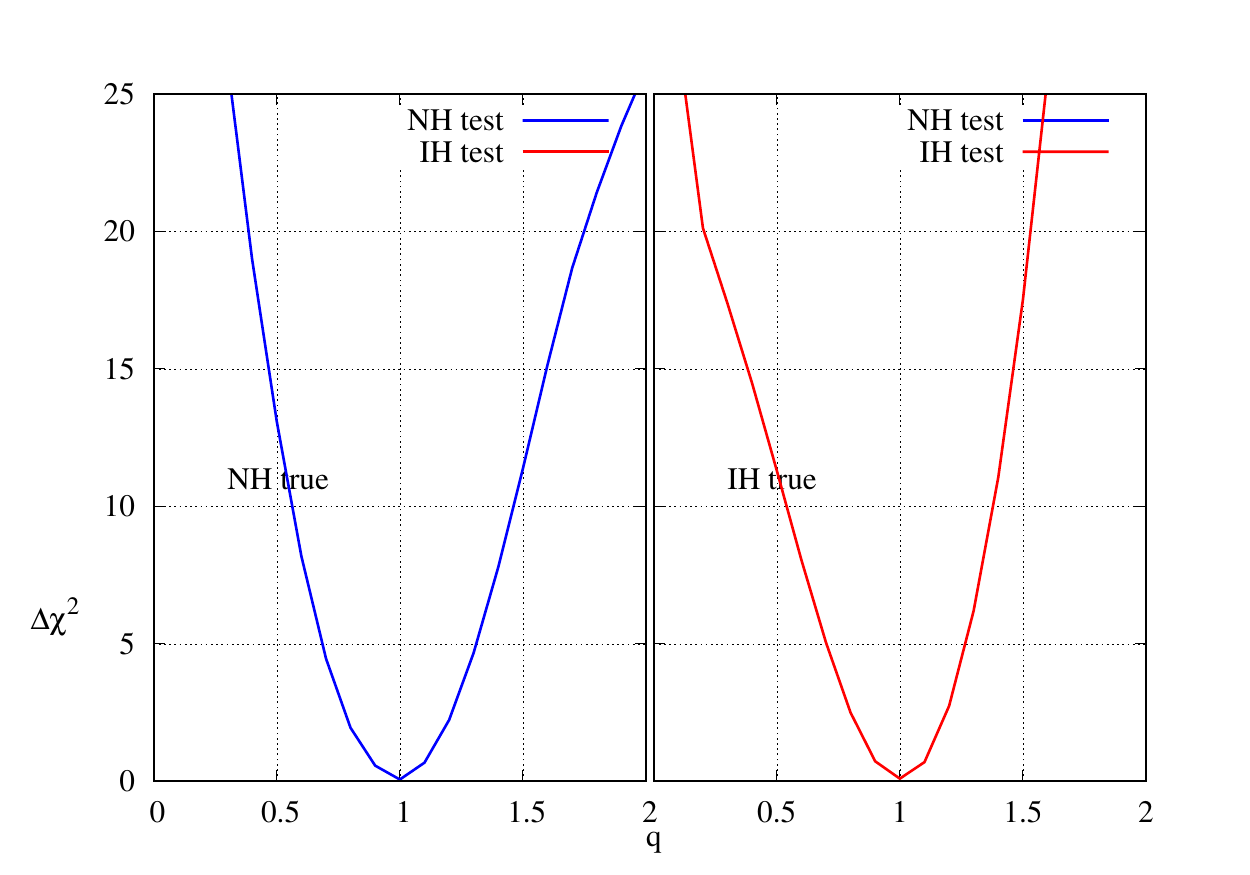}
\caption{\footnotesize{$\Delta \chi^2$ vs $q$ for an expected $(5 \nu+5 \bar{\nu})$ run of DUNE plus equal 
$\nu$ and $\bar\nu$ runs of T2K with $37.4\times 10^{20}$ POT and of NO$\nu$A with $30.25\times 10^{20}$ POT. 
The left (right) panel assumes the true hierarchy to be normal (inverted). The blue (red) curves assume the test 
hierarchy to be normal (inverted). 
}}
\label{fig4}
\end{figure}

The future long-baseline accelerator neutrino experiment DUNE~\cite{Abi:2018dnh,Abi:2018alz,Abi:2018rgm} is designed 
to disentangle the 
changes due to matter effects from the changes due to $\dcp$. Its baseline ($L\simeq1300$ km) is much longer than 
that of T2K or NO$\nu$A. Its peak energy is correspondingly higher and matter effects larger. Therefore, it is 
expected that it will have a much better ability to rule out vacuum oscillations. Fig.~\ref{fig3} shows the ability 
of one year neutrino run of DUNE to establish matter modified oscillations. We note that vacuum oscillations are 
ruled out at $3~\sigma$ if the true hierarchy is NH but only at $2~\sigma$ if the true hierarchy is IH. Addition of 
T2K ($5\nu +5\bar\nu$) and \nova ($5\nu +5\bar\nu$) runs leads only to a marginal improvement but not $3~\sigma$ 
discrimination. In order to consider a $5~\sigma$ discrimination against vacuum oscillations, we did a simulation of 
DUNE ($5\nu +5\bar\nu$) run. Here again, we find that a $5~\sigma$ discrimination is possible only if the true 
hierarchy is NH but not if the true hierarchy is IH. However, a $5~\sigma$ discrimination is possible for both 
hierarchies if the data of DUNE ($5\nu +5\bar\nu$) run is considered in conjunction with T2K ($5\nu +5\bar\nu$) 
and \nova ($5\nu +5\bar\nu$) runs, as illustrated in fig.~\ref{fig4}. We also note from this figure that values of 
$q$ out side the range ($1\pm 0.4$) are ruled out at $3~\sigma$ or better.     

\section{Conclusions}
Matter effects allow us to determine the sign of neutrino mass-squared differences. The existence of matter effects 
at the scale of $\ds$ is well established~\cite{Fogli:2005cq}. However, at the scale of $\dl$, vacuum oscillations  
fit the data nearly as well as matter modified oscillations. This is true for both atmospheric neutrino data~\cite{Abe:2017aap} 
and for present long-baseline accelerator data, as demonstrated in this work. We also show that even extended 
runs of T2K and \nova have {\bf no} discriminating ability against vacuum oscillations. A $3~\sigma$ discrimination 
against vacuum oscillations can be achieved with one year neutrino run of DUNE, if NH is the true hierarchy but not 
if IH is the true hierarchy. Ruling out vacuum oscillations at $5~\sigma$ requires the combined data of 
($5\nu + 5\bar\nu$) runs of T2K,  \nova and DUNE. Such a data can also establish the strength of matter effects 
with good precision.  

\section*{Acknowledgements}
SB thanks IIT Bombay for financial support.

\bibliographystyle{apsrev}
\bibliography{referenceslist}


\end{document}